# Strain-Induced Violation of Temperature Uniformity in Mesoscale Liquids

Eni Kume, Patrick Baroni, Laurence Noirez*

Lab. Léon Brillouin (CEA-CNRS), Univ. Paris-Saclay, CEA-Saclay, 91191 Gif-sur-Yvette Cédex, France

**Thermo-elasticity couples the deformation of an elastic (solid) body to its temperature and *vice-versa*. It is a solid-like property. Highlighting such property in liquids is a paradigm shift: it requires long-range collective interactions that are not considered in current liquid descriptions. The present pioneering microthermal studies provide evidence for such solid-like correlations. It is shown that ordinary liquids emit a modulated thermal signal when applying a low frequency (Hz) mechanical shear stress. The liquid splits in several tenths microns wide thermal bands, all varying synchronously and separately with the applied stress wave reaching a sizable amplitude of ± 0.2 °C. This thermal property is unknown in liquids. Thermo-mechanical coupling challenges a dogma in fluid dynamics: the liquid responds collectively, adapts its internal energy to external shear strain and is not governed by short relaxation times responsible of instant thermal dissipation. The proof of thermomechanical coupling opens the way to a new generation of energy-efficient temperature converters.**

Control and understanding of small-scale flows is rich in potential applications. Because macroscopic laws for flow and thermodynamics reach often the frontiers of their validity, the physics remains difficult but fascinating. Modern techniques give us the opportunity to probe more precisely physical properties at the sub-millimeter scale, to uncover new and major knowledges that challenge the concept of continuum mechanics. Low frequency shear elasticity belongs typically to these newly identified properties [1-7]. It tells that liquid molecules might be long-range elastically correlated and thus challenges the conventional definition of fluids that foresees shear elastic properties only at very high frequency; i.e. when the fluid excited at time scales greater than a molecular relaxation time defined as $\tau = \eta/G$, where $\eta$ is the viscosity and $G$ the shear elastic modulus. These high frequency excitations are typically in the megahertz – gigahertz range corresponding to shear elastic moduli of the order of MPa-GPa [8-10]. This commonly adopted approach is due to Frenkel (1950) [8]. In contrast, the low frequency shear elasticity concerns the Hertz domain, is weak of the order of 1-$10^3$ Pa, negligible in practise and depends on the scale and the fluid considered [5-7,11-14]. Recently, a revision of the Frenkel calculation has revealed the pertinence of considering the low frequency shear elasticity specifying the conditions under which it applies (k-gap model) [3,4].

On the assumption of low frequency liquid elasticity, it turns out the possibility to reveal novel liquid properties. This is the goal of the present study, which is pioneering for several aspects. To access physical properties of a material, a general principle is to submit it to an external stress and to observe its response function. Here, the liquid is submitted to a low frequency (~ Hz) mechanical shear strain (external stress) and we probe its dynamic state via accessing the instant temperature on each point of the liquid gap (scheme 1).

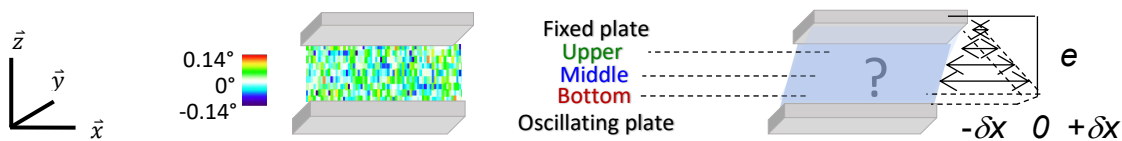

**Scheme 1**: The liquid is confined between two high energy plates (Alumina surfaces). At left, the thermal image of the liquid is viewed at rest from the gap plane (xOz) (63 x 13 pixels (renormalized) thermal image of glycerol

recorded at room temperature and 0.240mm gap). The impact of a low frequency (~ Hz) mechanical shear strain on the liquid temperature is studied.

In the low frequency conditions, the mechanical energy transmitted during an oscillation cycle is extremely weak and negligible (for glycerol: $\eta = 1$ Pa.s at the largest oscillation amplitude ($\gamma = 4000\%$) at $\omega = 1$ rad/s, the energy is about $10^{-3}$ Joule/cm$^3$, with a specific heat $c_p = 2450$ J.kg$^{-1}$.K$^{-1}$, the maximum temperature increase is less than $10^{-3}$°C). Additionally, a low frequency excitation is much slower than any known relaxation time [8-10], ruling out a dynamic coupling with a molecular relaxation time ($1/\omega \ll \tau_{relax}$) [15]. Likewise for thermal fluctuations, that scale with $\tau_{fluct} = \xi^2/D$ where, $\tau_{fluct}$ is the lifetime of the density fluctuations, $\xi$ being of the order of the nanometer and $D$ the diffusion coefficient, is of the order of $10^{-9}$ m$^2$/s [16].

The present study reveals instead the emergence of a modulated thermal signal upon applying a low frequency mechanical shear stress, which firmly contradicts of the above assumptions. The chosen liquids are the glycerol and the polypropylene glycol due to the negligible evaporation rate at room temperature (i.e. away from any phase transition). We identify local negative and positive temperatures spread in about 50µm width strata, alternating cold and warm zones, synchronously with the applied deformation. The temperature variations are significant (about $\Delta T \pm 0.2$°C) reproducible and reversible; they coexist without dissipating in the noise of the thermal fluctuations. The liquid accommodates locally its internal energy versus the external stress field. This process does not involve any relaxation time, but reveals a shear mechanical property of liquids.

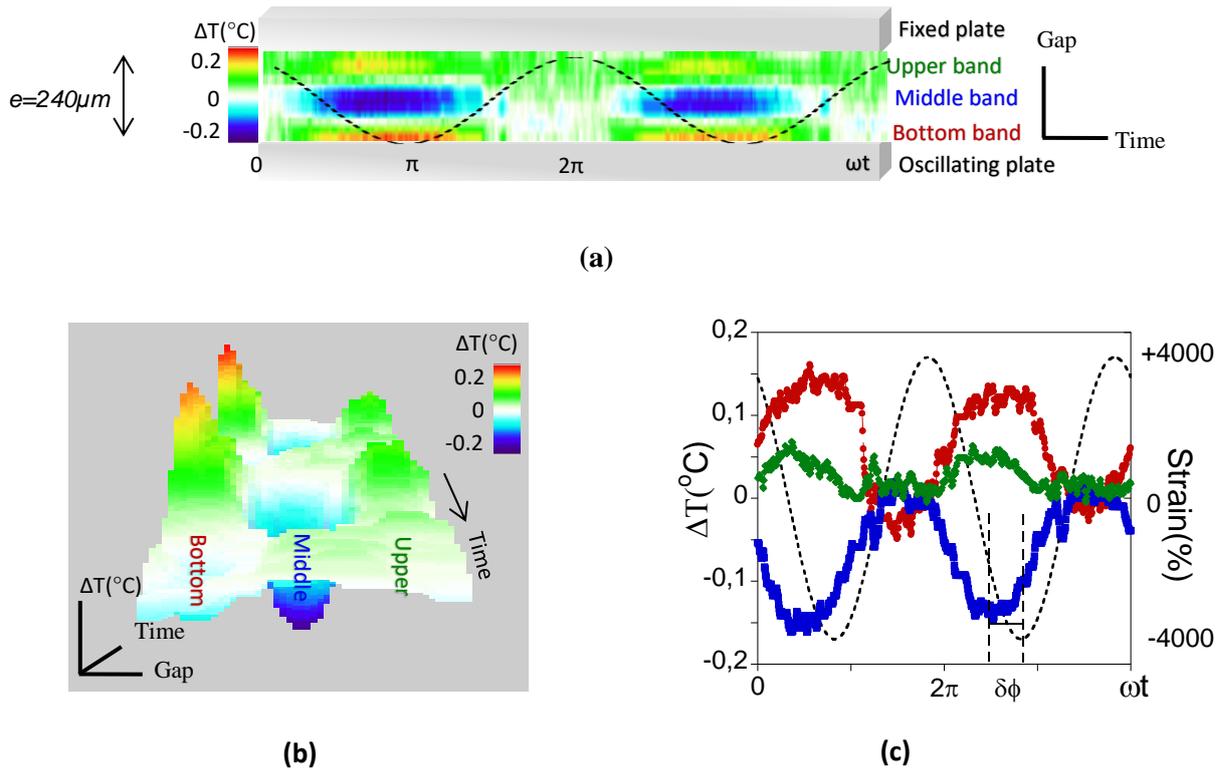

**Figure 1.** By applying a low frequency mechanical stimulus (~ Hz), the liquid emits a modulated thermal signal synchronous with the stimulus. **a)** Real-time mapping of the temperature of the PPG-4000 confined in a 240µm gap (gap view) excited with a low frequency oscillatory shear strain (room temperature measurements at $\omega=0.5$ rad/s and $\gamma = 4000\%$ gathering about ~800 frames, alumina substrate). Black line is an eye guide for the applied strain. **b)** 3D view of the thermal behaviour of Fig 1a. **c)** Temperature variations of the bottom, middle and upper bands displayed in red (●), blue (■) and green (◆) respectively. The dotted line is an eye guide for the applied shear strain. δφ is the phase shift between the thermal and strain wave. The zero $\Delta T$ is defined relative to the observation window.

Figures 1a-b display the thermal images of the low molecular weight polypropylene glycol (PPG-4000) during two periods ($\omega=0.5$rad/s) of the oscillatory shear strain (at $\gamma = 4000\%$). This representation overviews the instant state of the system at different stages of the mechanical deformation. The thermal image shows the periodic emergence of coexisting hot and cold zones distributed in three thermal bands. The bands exhibit opposite thermal behaviours. While the central band is cooling down, the neighboured ones are heating up. This leads to a temperature compensation in the liquid volume (Fig.1a and c). These thermal changes are in advance with the applied strain wave (Fig. 1c), the phase shift being approximately $\pi/4$ ($\delta\varphi_{middle} \sim 51° \pm 2°$, $\delta\varphi_{bottom} \sim 46° \pm 2.5°$ and $\delta\varphi_{upper} \sim 48.25° \pm 2°$, for the middle, bottom and upper bands respectively). From a mechanical point of view, reversible temperature changes induced by the deformation is known in solids as the thermo-(shear)elastic effect [17]. The applied slow dynamics exclude a coupling with time-dependent processes such as the viscoelastic relaxation time (for glycerol $\tau_{relax} \sim 10^{-8}$s) [18]. Consequently, the applied strain is the source of the thermal effect. This reversible thermal modulation rules out mechanisms related to thermal conduction and therefore may be regarded as adiabatic.

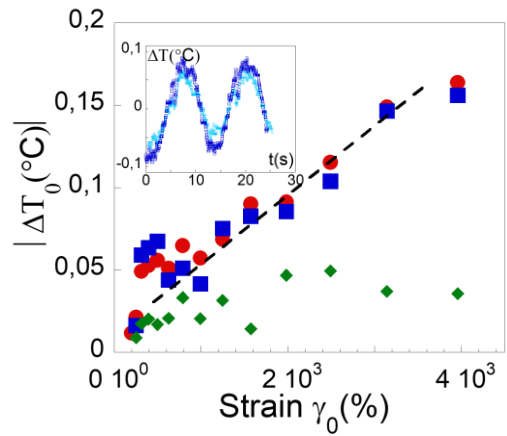

**Figure 2.** Strain dependence of the amplitude maximum of the temperature variation in absolute units $|\Delta T_0(°C)|$. Sample: PPG-4000 at gap thickness 0.240mm, $\omega=0.5$ rad/s, for bottom "hot" band: (●), middle "cold" band: (■) and upper band: (◆) respectively - measurements below 150% are below the accuracy. The insert illustrates the thermal waves at gap thickness for strain values 2500% (light blue points) and 4000% (blue points) of the middle band (0.240mm gap, $\omega=0.5$rad/s).

Figure 2 shows the strain dependence of the temperature variation (amplitude maximum in absolute units) in the three coexisting thermal bands identified in Fig.1. Cold and hot waves vary in opposite way with increasing strain amplitude (becoming "cooler" and "warmer" respectively). This representation highlights the superposition of "hot" and "cold" values in a nearly exact compensation confirming that the system works adiabatically (without energy external transfer). The inset of figure 2 illustrates the thermal waves at two different strain values (for the same band). The temperature variation is not measurable at low strain and increases linearly with the strain amplitude above $\gamma > 200\%$ reaching a maximum variation of ~0.2°C at 4000 % (10 times over the accuracy). The linear relation holds true for the two thermal bands of opposite behaviour closer to the moving surface, while the evolution for the band farthest from the moving surface is nearly strain-independent. The proportionality between the two quantities: temperature variation and mechanical deformation (at low strain) defines a thermo-elastic constant $K(\Delta T, \gamma)$. The linear relationship is a strong indication of a thermal mechanism occurring as soon as the smallest shear strain values; i.e. inherent to the liquid deformation.

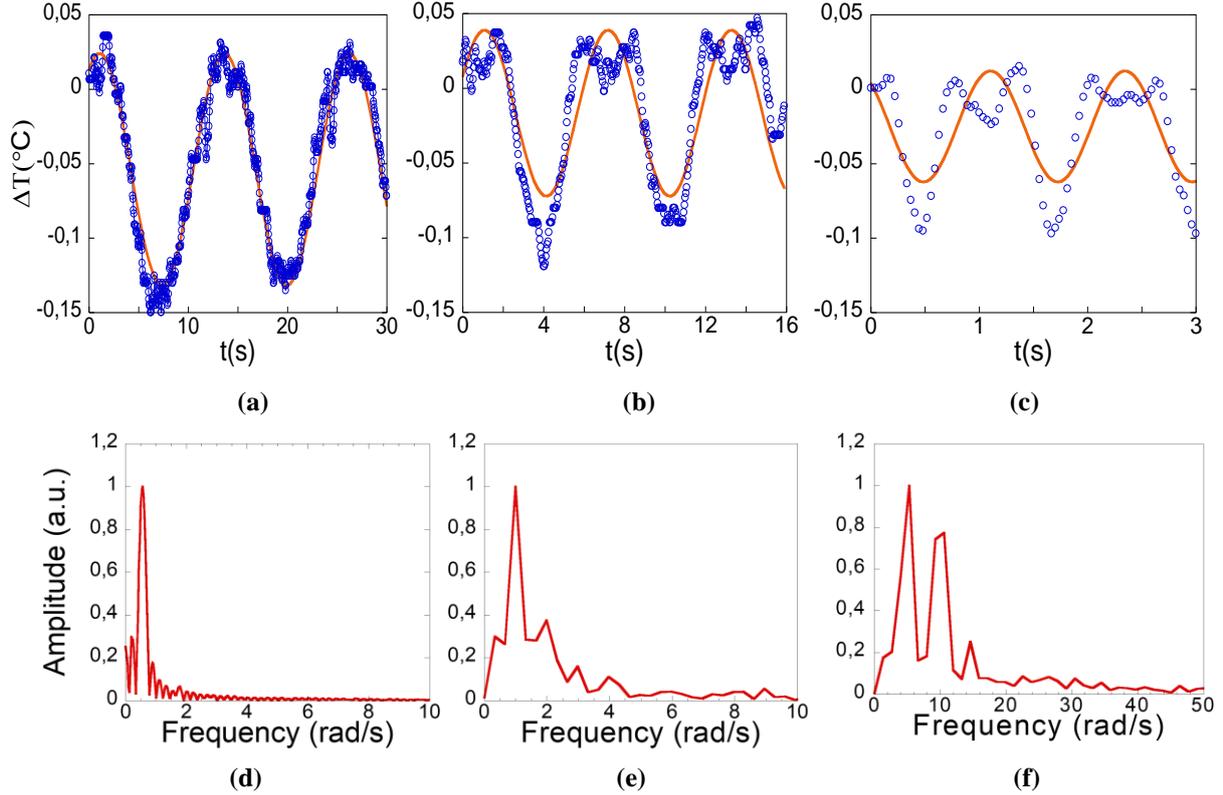

**Figure 3.** Data points (blue circles) and sin fit (continuous orange line) of the thermal variation ΔT within PPG-4000 middle band at 0.240mm, γ=4000%, **a)** ω=0.5 rad/s, **b)** ω=1 rad/s, **c)** ω=5 rad/s. Corresponding FFT analysis of the waves depicted at **d)** Fig. 3a, **e)** Fig. 3b, **f)** Fig. 3c. The zero ΔT is relative to the considered (middle) band.

The impact of the frequency on the thermal behaviour is investigated in a range of 0.5 - 5 rad/s and strain values ranging from 0.5% - 4000%. At low frequency, the thermal variation reproduces the input strain waveform (Fig.3a recorded at ω=0.5rad/s). As the frequency is increased, the thermal response progressively loses the input sinus shape (figures 3b, 3c) and the amplitude of thermal variation is decreasing of nearly 40% from 0.5 to 5 rad/s, implying the excitation frequency is faster than the relaxation between two successive oscillations. A FFT analysis (Fig. 3d-3f) shows that in addition to the fundamental frequency (Fig.3d), higher harmonics emerge such as, the second and third harmonics at high frequencies (Fig 3b, 3c). The generation of thermal harmonics might be due to non-linear elastic effects, fluid inertia [19,20], slippage instabilities, backflow and viscous friction heating [21].

The glycerol exhibits a similar thermal behaviour upon applying shear strain stress (Fig. 5a), but the thermal variation is weaker compared to PPG-4000 (ΔT$_{max}$~0.12°C). This weaker response can be interpreted by the key role of the total entropy (intermolecular, conformational). Glycerol is a molecular liquid with symmetrical geometry whereas the low molecular weight PPG-4000 is an oligomer, meaning increased conformational entropy and multiple intermolecular interactions per molecule. Thermal bands are identified with varying gaps from 45μm. More precisely three thermal bands seems to be the rule for various gap thickness between 150μm - 500μm. It is systematically observed that the bands oscillate in opposite temperature variation implying that the temperature compensation is essential for the mechanism. The spatial position of each band is not fixed along the gap thickness. Above 500μm, multiple bands (> 3) appear hardly observable being "diluted" in the width of the gap (Fig.5b). The thermal mapping (Fig. 1a, Fig. 5b), the temperature graph (Fig.1c) and the strain dependence graph (Fig. 2) indicate that the upper band (farthest band from the moving surface) exhibits the weaker thermal signal with respect to the other ones. The thermo-mechanical effect primarily takes place near the motion surface (strain source) which means that the energy might not necessarily propagate along the whole gap.

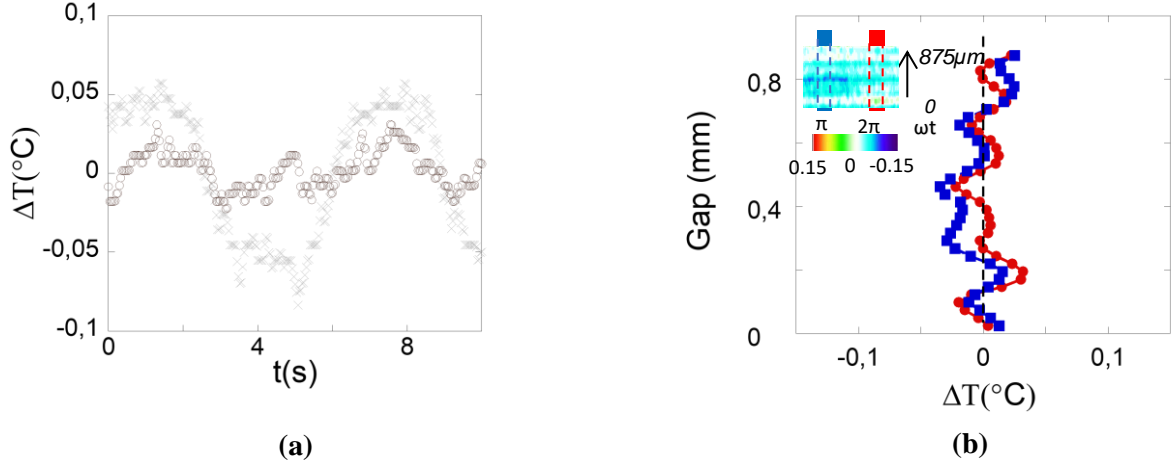

**Figure 5. a)** Comparison of Glycerol and PPG-4000: Brown circles (○) and grey crosses represent the temperature variations of glycerol and PPG-4000 on one (bottom) thermal band at 0.240mm gap thickness, ω =1rad/s and γ =2500% respectively. **b)** Vertical cut of the thermal changes (°C) in glycerol under oscillatory shear strain at large gap thickness (0.875mm), ω = 1rad/s and γ = 650 % (room temperature measurements). Red and blue data correspond to the temperature along the gap in the areas marked in red and blue in the inset (~400 frames).

How to understand that viscous liquids convert a low frequency shear strain in synchronous thermal waves. Such a collective property rules out a viscoelastic coupling with relaxation times ($1/\omega \ll \tau_{relax}$). Furukawa and Tanaka [22] suggest that liquids can be compressible by shear flow in order to explain the cavitation creation. They base their study on the density fluctuations. The theory "does not involve any viscoelastic or structural relaxation" [22]. However, this interpretation is here not satisfying since the expected shear rate is about $10^6$ s$^{-1}$, much higher than the inverse of the timescales corresponding to the low frequency range explored in this study.

A mechanism alternating compressed-stretched states in volume and time seems plausible; the "stretched" state generates the cold area while the "compressed" state generates the hot. These two thermodynamic states compensate dynamically: i.e. the system does not exchange with the environment ($dQ = 0$). The liquid stores the shear wave energy to self-organize in local temperature bands without energy transfer and therefore this adiabatic process can repeat endlessly. We also showed that the thermal signal increases linearly with the strain amplitude for slow dynamics (ω=0.5rad/s) highlighting a thermoelastic behaviour that is usually expected in solids only and generates harmonics at high frequency. In solids upon applying a periodic strain (load) [17, 23], periodic thermal changes exhibit temperature oscillations in opposite phase with the applied strain [17]. A similar behaviour is noticed for the (bottom) band of the liquid near the moving plate. The self-organisation of the liquid in several temperature bands working with opposite temperature variations of few tenths of a degree might be understood as a consequence to minimize the increase of internal energy due to the strain. A comparison with the temperature variation $\Delta T$ in solid thermoelastic processes can be made: $\Delta T = -T_0 \frac{\alpha}{\rho c_p} \Delta \sigma$ with $\alpha$ the linear expansion coefficient, $\rho$ the density of the material, $T_0$ the reference temperature and $c_p$ the specific heat at constant pressure [17]. The temperature variation is linear with the applied stress $\Delta \sigma$ in agreement with the strain dependence observed for the liquid PPG at moderate strain values (Fig.2). However, the amplitude of the thermal wave at solid surfaces is much weaker than the liquid (0.001°C against 0.2°C) and is hardly comparable [17].

The thermal mapping reveals the coexistence of several thermal states; i.e. the stress is not uniformly distributed within the gap. It is important to outline that thermal local states are visible at high strain rates only; i.e. in highly non-equilibrium conditions for which the Second Law of thermodynamics does not apply. The ability to convert the shear deformation energy in thermal states, implies that density (thermal) fluctuations and liquid molecules are long-range elastically correlated. Under local potential (stress, surface vicinity, etc), correlated systems typically do not relax to the equilibrium. It evidences that mechanical properties of liquids must be rather treated as long range shear-elastically correlated systems that reach nearly stable non-equilibrium states [24], where the injected mechanical energy can be adiabatically transformed generating heating and cooling [25]. Frenkel [8] was one of the first to propose the existence of shear (transverse) modes in liquids. Above $\omega > \omega F$, where $\omega F$ is Frenkel frequency, liquids would support both longitudinal and transverse waves. However, the relaxation time (called Maxwell relaxation time $\tau M$) is much faster than the applied (low frequency) mechanical excitations. Trachenko et al. [3, 4, 26] revisited the Frenkel theory. They use a generalized viscosity that contains short-term elasticity and get a new frequency $\omega = \sqrt{c^2 k^2 - \frac{1}{4\tau^2}}$ [3, 4, 26], where propagation modes exist and show the existence of k-gap. As mentioned in [4], "from the point of view of elasticity, the k-gap suggests that we can consider a liquid as collection of dynamical regions of characteristic size cτ where the solid-like ability to support shear waves operate". This could imply that $\tau$, "the time between consecutive diffusive jumps in the liquid" [4,8] or shear wave velocity c, is scale-dependent, in agreement with the present mesoscopic thermal observations. The existence of a dynamic "network" is also supported by the recent identification of liquid shear elasticity [5, 11-14]. As a result, when the molecules are dynamically correlated, the thermodynamic variables pressure, volume, temperature and entropy are no longer independent [27]. The correlated liquid molecules nearly instantaneously adapt the thermodynamic state according to the applied strain. This thermal effect is identified far away from any critical point (here the glass transition), suspecting a generic thermo-mechanical property. These are new insights for the microfluidic understanding. Important applications for thermal control in micro-devices and heat sinks can be foreseen using and controlling these newly uncovered thermo-mechanical liquid properties.

## Acknowledgements

This project has received funding from the European Union's Horizon 2020 research and innovation programme under the Marie Sklodowska-Curie grant agreement N° 766007.

**Author contribution:** E.K. undertook the experimental work and wrote the paper. P.B. was responsible of the instrumental setting and protocol of image analysis. L.N. proposed the scientific subject and supervised the developments.

## Methods

The thermal emission was recorded using an infrared sensor equipped with a macro-lens focussing the liquid gap in between the two plates. The frame rate was 27 Hz – and the depth of field (DOF) is ~0.85mm. The thermal accuracy is ± 0.020°C. For temperatures near ambient temperature, the emitted radiation is in the near infrared range (700nm – 1000μm). Based on the Stefan-Boltzmann law:

E= $\varepsilon_m \sigma (T^4 - T_c^4)$ where $E$ is the energy flux, $\varepsilon_m$ the emissivity, $\sigma$ the Stefan–Boltzmann constant, $T$ the temperature and $T_c$ the environment temperature. All the presented results have been corrected by subtracting the median value measured at rest prior the dynamic measurements. An example of thermal mapping of liquid at rest is displayed at scheme 1. The 2D mapping versus time were obtained by construction of a kinetic image made of the succession of each frame (1 frame/0.037s).

The liquid was confined between two disk-like α-Alumina high-energy surfaces to increase the interaction between the surface and the liquid molecules [12-13]. The surfaces have been heat treated beforehand to remove any organic component and moisture. This procedure ensures the total wetting of the liquid. Alumina has a relatively low thermal conductivity (28 - 35 W/m$^{-1}$K$^{-1}$ at room temperature) compared to metallic substrates (Aluminum ~230 W/m$^{-1}$K$^{-1}$). The liquid which is itself poorly thermally conductive (0.30 W/m$^-$1K$^{-1}$). The geometry of our setup is a conventional plate-plate one. Shear strain is applied by oscillating the bottom surface (Fig.1). The upper surface is coupled with a sensor, which measures the shear stress (torque) transmitted by the liquid. Keithley multimeters record strain and stress mechanical waves with high accuracy. Both mechanical and thermal measurements were recorded simultaneously.

The liquids, glycerol and polypropylene glycol (PPG-4000, $M_n$=4000) are nearly black bodies in mid- and near-infrared radiations. Mechanical and thermal behaviours were systematically probed for gap thickness from 500μm down to 85μm and at a frequency range from 0.5 to 5 rad/s, at room temperature away from any critical point (glass transition temperatures for glycerol and PPG are - 93°C and - 73°C respectively).